\documentclass[twocolumn,showpacs]{revtex4}
\usepackage{graphicx}

\def\l{\langle}
\def\r{\rangle}
\newcommand{\vecS}{\mbox{\boldmath $S$}}

\begin{document}

\title{
Finite-size Scaling of Correlation Ratio and 
Generalized Scheme \\ for the Probability-Changing Cluster Algorithm}

\author{Yusuke Tomita}
\email{ytomita@phys.metro-u.ac.jp}
\affiliation{
Department of Physics, Tokyo Metropolitan University,
Hachioji, Tokyo 192-0397, Japan
}

\author{Yutaka Okabe}
\email{okabe@phys.metro-u.ac.jp}
\affiliation{
Department of Physics, Tokyo Metropolitan University,
Hachioji, Tokyo 192-0397, Japan
}

\date{\today}

\begin{abstract}
We study the finite-size scaling (FSS) property of the correlation ratio, 
the ratio of the correlation functions with different distances. 
It is shown that the correlation ratio is a good estimator 
to determine the critical point of the second-order transition 
using the FSS analysis.  The correlation ratio is especially useful 
for the analysis of the Kosterlitz-Thouless (KT) transition. 
We also present a generalized scheme of 
the probability-changing cluster algorithm, which has been 
recently developed by the present authors, based on the FSS property 
of the correlation ratio.  
We investigate the two-dimensional quantum 
XY model of spin 1/2 with this generalized scheme, 
obtaining the precise estimate of the KT transition temperature 
with less numerical effort.  
\end{abstract}


\maketitle

The development of efficient Monte Carlo algorithms is 
important for studying many-body problems in physics.  
We recently developed a cluster algorithm,
which is called the probability-changing cluster (PCC) algorithm,
of locating the critical point automatically \cite{PCC}.  
It is an extension of the Swendsen-Wang (SW) algorithm \cite{SwWa}, 
but we change the probability of cluster update 
(essentially, the temperature) during the Monte Carlo process.  
Since we do not have to make simulations for several parameters, 
we can extract information on critical phenomena with much 
less numerical effort.  We applied the PCC algorithm to the 2D 
diluted Ising model \cite{dil2d}, investigating the crossover 
and self-averaging properties. 
We also extended the PCC algorithm to the problem of 
the vector order parameter \cite{XY} with the use of Wolff's embedded 
cluster formalism \cite{Wolff89}; studying the 2D classical 
XY and clock models, we showed that the PCC algorithm 
is also useful for the Kosterlitz-Thouless (KT) transition \cite{KT}.

In the original formulation of the PCC algorithm \cite{PCC}, 
we used the cluster representation of the Ising model (generally, 
the Potts model) due to Kasteleyn and Fortuin 
\cite{KF} in two ways.  First, we make a cluster flip as 
in the SW algorithm \cite{SwWa}. 
Second, we change the probability of connecting spins of 
the same type, $p$, depending on the observation 
whether clusters are percolating or not. 
We use the finite-size scaling (FSS) relation for 
the probability that the system percolates, $E_p$, 
\begin{equation}
  E_p(p,L) = X(t L^{1/\nu}), \quad t=(p_c-p)/p_c ,
\label{scale}
\end{equation}
to determine the critical point. 
Here, $L$ is the system size, $p_c$ is the critical value of $p$ 
for the infinite system, and $\nu$ is 
the correlation-length critical exponent. 
We may alternatively consider that $t=(T-T_c)/T_c$, where $T$ 
is the temperature.  
With a negative feedback mechanism, we locate 
the size-dependent temperature, $E_p$ of which is 1/2. 
The point is that $E_p$ has 
the FSS property with a single scaling variable. 
We may use quantities other than $E_p$ which have 
a similar FSS relation.
Then, we could generalize the PCC algorithm 
for a problem where the mapping to the cluster
formalism does {\it not} exist.  

In the FSS analysis of the simulation, 
we often use the Binder ratio \cite{Binder}, which is 
essentially the ratio of the moments of the order parameter $m$. 
The moment ratio has the FSS property with a single scaling variable,
\begin{equation}
  \frac{\l m^4 \r}{\l m^2 \r^2} = f(tL^{1/\nu}),
\label{mom_ratio}
\end{equation}
as far as the corrections to FSS are negligible.  Here, the angular 
brackets indicate the thermal average.  
However, the moment ratio derived from a snapshot spin configuration 
is always one; therefore, the instantaneous moment ratio 
cannot be used for the criterion 
of judgment whether we increase or decrease the temperature. 

In this Letter, we treat the correlation ratio, 
the ratio of the correlation functions with different distances. 
Studying the thermodynamic properties of the correlation ratio, 
we will show that it is a good estimator 
for determining the critical points of the second-order and 
KT transitions.  
We also propose a generalized scheme of 
the PCC algorithm based on the correlation ratio.  
Using this general scheme combined with 
the quantum Monte Carlo simulation 
of the continuous-time loop algorithm, 
we study the 2D quantum XY model of spin 1/2. 

Let us start with considering the spin-spin correlation 
for the $D$ dimensional systems, 
\begin{equation}
 g(r) = \frac{1}{N} \sum_{i} \vecS_i \cdot \vecS_{i+r},
\end{equation}
where $\vecS_i$ is a spin at site $i$, and 
$N=L^D$ is the number of spins.  Precisely, the distance $r$ is 
a vector, but we have used a simplified notation.  
We have assumed a translational invariance. 
For an infinite system at the critical point, the correlation 
function decays as a power of $r$, 
\begin{equation}
  \l g(r) \r \sim r^{-(D-2+\eta)}, \quad (L=\infty, \ t=0),
\label{g(r)}
\end{equation}
with the decay exponent $\eta$. 
Away from the critical point, the ratio of the distance $r$ 
and the correlation length $\xi$ plays a role 
in the scaling of the correlation function. 
Moreover, for finite systems, two length ratios come in 
the scaling form of the correlation function;
\begin{equation}
 \l g(r,t,L) \r
 \sim r^{-(D-2+\eta)} \ h(r/L, L/\xi), \quad
 (L \ne \infty, \ t \ne 0). 
\end{equation}
Then, the ratio of the correlation functions with 
different distances $r$ and $r'$ becomes 
\begin{equation}
 \frac{\l g(r,t,L) \r}{\l g(r',t,L) \r} = 
 \Big( \frac{r}{r'} \Big)^{-(D-2+\eta)} \ 
 \frac{h(r/L, L/\xi)}{h(r'/L, L/\xi)}.
\end{equation}
If we fix two ratios, $r/L$ and $r/r'$, the correlation ratio 
takes the FSS form with a single scaling variable, 
\begin{equation}
 \frac{\l g(r,t,L) \r}{\l g(r',t,L) \r}= \tilde f(L/\xi).
\label{corr_ratio}
\end{equation}
In case the correlation length diverges with a power law, 
$\xi \propto t^{-\nu}$, Eq.~(\ref{corr_ratio}) becomes 
the same form as Eq.~(\ref{mom_ratio}); however, we should 
note that Eq.~(\ref{corr_ratio}) is also applicable 
to the case of the KT transition, where the correlation 
length diverges more strongly than the power-law divergence. 

\begin{figure}
\includegraphics[width=0.89\linewidth]{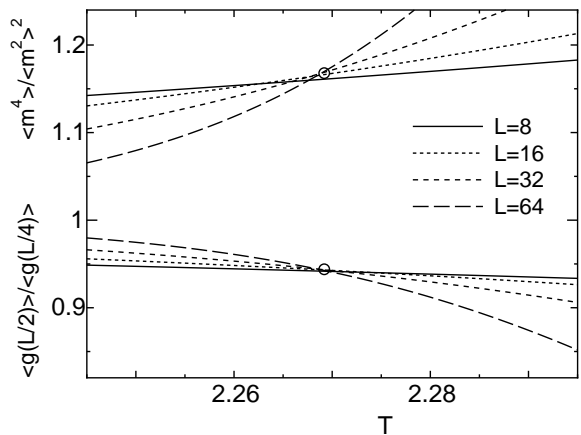}
\caption{Plot of the moment ratio $\l m^4 \r/\l m^2 \r^2$ and 
the correlation ratio $\l g(L/2)\r/\l g(L/4) \r$
of the 2D Ising model for $L$ = 8, 16, 32, and 64. 
The critical values of two ratios for $L=\infty$ are given 
by open circles. 
} 
\label{fig_1}
\end{figure}

In order to examine the FSS properties of the correlation ratio, 
we simulate the 2D Ising model on the square lattice 
with the periodic boundary conditions. 
As for the distances $r$ and $r'$, we choose $L/2$ and $L/4$; 
we take the horizontal or vertical direction of the lattice 
for the orientation of two sites. 
We show the temperature 
dependence of the moment ratio $\l m^4 \r/\l m^2 \r^2$ 
and the correlation ratio $\l g(L/2) \r/\l g(L/4) \r$ 
in Fig.~\ref{fig_1}. 
From now on, we represent the temperature in units of $J/k_B$, 
where $J$ is the coupling constant and $k_B$ is 
the Boltzmann constant. 
The error bars are the order of the width of curves. 
We see from Fig.~\ref{fig_1} that the crossing of the data 
of different sizes are better for the correlation ratio; 
in other words, the corrections to FSS for the correlation ratio 
are smaller than those for the moment ratio.  
We check the two ratios at the critical point 
for a quantitative comparison. 
The size dependence of the moment ratio for the 2D Ising model 
was carefully studied by Salas and Sokal \cite{Salas}. 
The correlation functions of the critical 2D Ising model 
for $L \to \infty$ can be calculated 
with the use of the continuum field theory \cite{Francesco}.
The moment ratio is written in terms of the integral of the 
$2n$-point correlation functions, and the fourth-order 
moment ratio becomes 
\begin{equation}
 \l m^4 \r/\l m^2 \r^2 = 1.1679229(47) 
\label{mom_cr}
\end{equation}
in the limit $L \to \infty$ \cite{Salas}, 
where the numbers in the parentheses denote the uncertainty 
in the last digits.  Without performing an integral, 
the correlation ratio is simply written as 
\begin{equation}
 \frac{\l g(L/2) \r}{\l g(L/4) \r} = 
 \frac{|\theta_1(1/2)|^{-1/4} \ \sum_{\nu=1}^4 |\theta_{\nu}(1/4)|}
      {|\theta_1(1/4)|^{-1/4} \ \sum_{\nu=1}^4 |\theta_{\nu}(1/8)|},
\label{theta}
\end{equation}
where $\theta_{\nu}(z)$ are the Jacobi $\theta$-functions.  
Thus, we can easily obtain the correlation ratio at the critical point as
\begin{equation}
 \l g(L/2) \r/\l g(L/4) \r = 0.943904982
\label{cor_cr}
\end{equation}
in the limit $L \to \infty$. 
The critical values of two ratios for $L=\infty$, 
Eqs.~(\ref{mom_cr}) and (\ref{cor_cr}), are shown 
by open circles in Fig.~\ref{fig_1}. 
The calculated correlation ratio at 
$T_c = 2/\ln(1+\sqrt{2})=2.269\cdots$  for $L=8$ is 0.9418(5). 
The deviation from the value for the infinite lattice is 
only 0.2\% even for $L=8$, 
which is smaller than that for the moment ratio, 0.7\%.  
We mention that the scaling plot of the correlation 
ratio as a function of $(T-T_c) L^{1/\nu}$ 
with $\nu=1$ is very good. 
Our results suggest that the correlation ratio 
is a good estimator to determine $T_c$ 
using the FSS analysis. 

\begin{figure}
\includegraphics[width=0.89\linewidth]{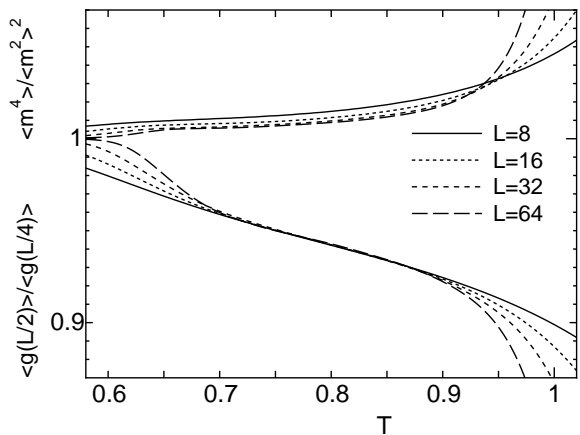}
\caption{Plot of the moment ratio $\l m^4 \r/\l m^2 \r^2$ and 
the correlation ratio $\l g(L/2) \r/\l g(L/4) \r$
of the 2D 6-state clock model for $L$ = 8, 16, 32, and 64. 
} 
\label{fig_2}
\end{figure}

Let us turn to the KT transition.  As an example, we pick up 
the 2D 6-state clock model. 
It was shown that the 2D $q$-state clock model has 
two phase transitions of the KT type at $T_1$ and $T_2$ 
($T_1<T_2$) for $q>4$ \cite{Jose}. 
We here simulate the 2D 6-state clock model on the square lattice. 
We show the temperature dependence 
of the moment and correlation ratios 
in Fig.~\ref{fig_2}.  For the correlation ratio, the curves 
of different sizes overlap in the intermediate KT phase 
($T_1<T<T_2$), and spray out for the low-temperature ordered 
and high-temperature disordered phases, which is expected from 
the FSS form of Eq.~(\ref{corr_ratio}). 
Then, we can make a FSS analysis based on the KT form of 
the correlation length, 
\begin{equation}
 \xi \propto \exp(c/\sqrt{t}).
\label{corr_length} 
\end{equation}
We try scaling plots of $\l g(L/2) \r/\l g(L/4) \r$ 
as a function of $L/\exp(c_1/\sqrt{1-T/T_1})$ and 
$L/\exp(c_2/\sqrt{T/T_2-1})$ for the low- and 
high-temperature sides, respectively. 
Using the data for $L$=8, 12, 16, 24, 32, 48, and 64, 
we estimate two KT transition temperatures. 
The best-fitted estimates are 
$T_1$=0.6980(22) and $T_2$=0.9008(14), which are compatible 
with the recent results using the PCC algorithm \cite{XY}, 
$T_1$=0.7014(11) and $T_2$=0.9008(6). 
On the contrary, as seen from Fig.~\ref{fig_2}, 
the corrections to FSS are larger for the moment ratio, 
which makes the FSS analysis difficult. 
We have shown that the correlation ratio is a good 
estimator especially for the KT transition. 
It is due to the fact that we only use the property of 
correlation function; the characteristic of the KT transition 
is that the correlation function shows a power-law decay 
at all the temperatures of the KT phase. 
We will discuss this point later. 

We may use the FSS properties of the correlation ratio for 
the generalization of the PCC algorithm.  
Instead of checking whether the clusters are percolating or not, 
we ask whether the instantaneous correlation ratio $g(L/2)/g(L/4)$ 
is larger or smaller than some fixed value $R_c$. 
Of course, we can use other sets of distances. 
We decrease (increase) the temperature, if $g(L/2)/g(L/4)$ 
is smaller (larger) than $R_c$. 
We start the simulation with some temperature. 
We make the amount of the change of temperature, $\Delta T$, 
smaller during the simulation; in the limit of $\Delta T \to 0$, 
the system approaches the canonical ensemble.  
In most cases near the critical temperature, $g(L/2)$ and 
$g(L/4)$ take some positive values for ferromagnetic systems.  
It sometimes happens 
that $g(L/2)$ and/or $g(L/4)$ becomes 0 or negative. 
We simply regard such a case that the system is disordered, 
and then increase the temperature. 

Here, we apply the generalized scheme of the PCC algorithm 
to the study of the quantum spin system.  
The loop algorithm, especially the continuous-time loop 
algorithm \cite{Beard}, has been successfully used for 
various quantum spin systems.  We deal with the 2D $S=1/2$ 
quantum XY model as an example, and determine 
the critical point automatically.  
The Hamiltonian is written as 
\begin{equation}
 {\cal H} = -J \sum_{\left<i,j\right>} (\hat S_i^x \hat S_j^x 
 + \hat S_i^y \hat S_j^y),  
\label{H_XY}
\end{equation}
where the spin operators $\hat S^{x,y}$ are one half of 
the Pauli matrices $\sigma^{x,y}$. 
We are concerned with the correlation 
\begin{equation}
 g(r) = \frac{1}{N} \sum_i (\hat S_i^x \hat S_{i+r}^x
                          + \hat S_i^y \hat S_{i+r}^y), 
\end{equation}
which is an off-diagonal element if we take $z$ axis 
as a quantized axis.  Although off-diagonal elements are 
difficult to calculate with the conventional quantum 
Monte Carlo simulation methods, they are calculated easily 
with the continuous-time loop algorithm \cite{Brower}. 
We should note that when calculating the correlation with 
the improved estimator, the correlation becomes non-negative. 

We treat the systems with linear sizes $L$ = 8, 12, 16, 24, 32, 
48, 64, 96, and 128. We start with $\Delta T$ = 0.005, 
and gradually decrease $\Delta T$ to the final value, 0.0001. 
After 20,000 Monte Carlo sweeps of determining $T_c(L)$, 
we make 10,000 Monte Carlo sweeps to take thermal average; 
we make 100 runs for each size to get better statistics 
and to evaluate the statistical errors. 
We calculate $g(L/2)/g(L/4)$ to check whether it is 
larger than $R_c$ or not; 
the value of $R_c$ is set to be 0.8. 

\begin{figure}
\includegraphics[width=0.9\linewidth]{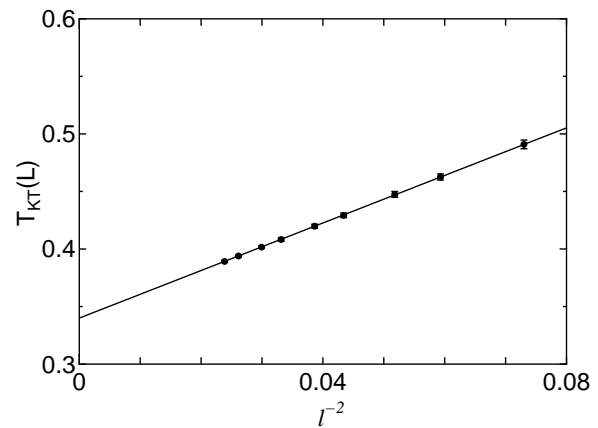}
\caption{Plot of $T_{KT}(L)$ 
of the 2D quantum XY model with $S$=1/2 for $L$ = 
8, 12, 16, 24, 32, 48, 64, 96, and 128, 
where $l = \ln bL$. 
} 
\label{fig_3}
\end{figure}

We follow the same procedure for analyzing the KT transition 
as the 2D classical XY model \cite{XY}. 
Using the FSS form of the correlation ratio and KT form of 
the correlation length (Eqs.~(\ref{corr_ratio}) and 
(\ref{corr_length})), we have the relation
\begin{equation}
 T_{\rm KT}(L) = T_{\rm KT} + \frac{c^2T_{\rm KT}}{(\ln bL)^2}.
\label{T_KT}
\end{equation}
We plot $T_{\rm KT}(L)$ as a function of $l^{-2}$ 
with $l=\ln bL$ for the best-fitted parameters in Fig.~\ref{fig_3}. 
The error bars are smaller than the size of marks. 
Our estimate of $T_{\rm KT}$ is 0.340(1). 
We estimate the uncertainty by the $\chi^2$ test of the data 
for 100 samples.  This value is compatible with the estimate 
of the recent study, 0.3427(2) \cite{Harada}. 
The constant $c$, in Eq.~(\ref{T_KT}), is estimated as $c$ = 2.45(5). 

\begin{figure}
\includegraphics[width=0.9\linewidth]{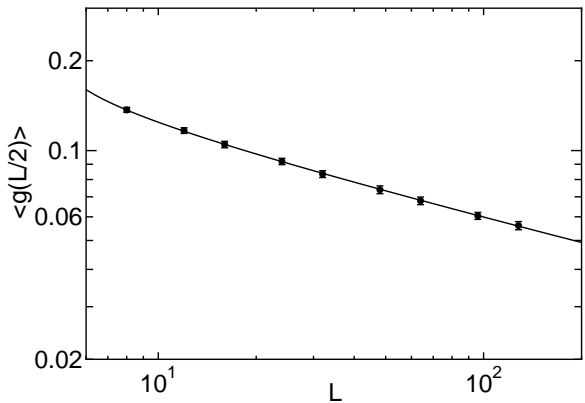}
\caption{Logarithmic plot of $\left< g(L/2) \right>$ 
at $T_{KT}(L)$ of the 2D quantum XY model with $S$=1/2 
for $L$ = 8, 12, 16, 24, 32, 48, 64, 96, and 128. 
} 
\label{fig_4}
\end{figure}

To discuss the critical exponent $\eta$, 
let us consider the correlation $\l g(L/2) \r$ at $T_{\rm KT}(L)$. 
In Fig.~\ref{fig_4}, we plot $\l g(L/2) \r$ as a function 
of $L$ in a logarithmic scale. 
In order to estimate $\eta$, we use the FSS form 
including small multiplicative logarithmic corrections 
\begin{equation}
 \l g(L/2) \r = AL^{-\eta}(\ln b^{\prime}L)^{-2r}; 
\label{eta}
\end{equation}
the existence of the multiplicative logarithmic 
corrections was pointed out for the KT transition 
\cite{Kosterlitz,Janke97}.  We should note that 
the logarithmic corrections are negligible 
for the correlation ratio. 
Using the above form, we obtain $\eta$ = 0.251(5) and $r$ = 0.055(9). 
We show the fitting curve obtained by using Eq.~(\ref{eta}) 
in Fig.~\ref{fig_4}. 
This value of $\eta$ is consistent with the theoretical prediction, 
1/4 (=0.25).  Our logarithmic-correction exponent $r$ takes 
almost the same value as that for the classical XY model; 
0.038(5) \cite{XY} and 0.0560(17) \cite{Janke97}. 

It should be noted that Alet and S{\o}rensen \cite{Alet} also 
applied the PCC algorithm to the quantum Monte Carlo simulation.  
One checks the ratio of the correlation length and the system size 
in their method.  

In summarizing, we have shown that the correlation ratio 
is a good estimator to determine the critical points of 
both the second-order and the KT transitions.  
A generalized scheme of the PCC algorithm based on the FSS property 
of the correlation ratio has been also presented.  

Although the study of the correlation ratio was partly 
motivated by a possible application to an efficient Monte Carlo algorithm, 
the FSS analysis of the correlation ratio itself is of great interest. 
The moment ratio \cite{Binder} has been employed in the study 
of various problems of the phase transition, 
from the lattice gauge theory to the polymer statistics.  
The use of the correlation ratio may help precise studies 
of the critical properties for complicated systems. 
The FSS analysis for the KT transition sometimes faces 
a difficulty due to the singular divergence of the correlation length 
and the logarithmic corrections \cite{Nomura95}, and may lead to 
false conclusions.  As a way to escape from such a difficulty, 
the level-spectroscopy method \cite{Nomura94} was developed 
for treating quantum systems which show the KT transition.  
Nomura \cite{Nomura95} studied the 
renormalization of the correlation functions for the 2D sine-Gordon model, 
and elucidated the logarithmic corrections on the fixed lines.  
It can be shown that the leading corrections are canceled if we consider 
the ratio of the correlation.  This is the reason why 
the correlation ratio is a very good estimator 
especially for the KT transition, which has not been noticed before. 
In contrast, the corrections may not be canceled for the moment ratio. 
In Fig.~\ref{fig_2} we have given a direct demonstration of 
the intermediate KT phase of the 2D clock model with rather small sizes, 
which clearly indicates the superiority of the correlation ratio 
in the analysis of the KT transition. 
The FSS of the correlation ratio is very promising for the study of 
various systems which show the KT transition. 

Our generalized scheme of the PCC algorithm is now free from 
the restriction of the cluster mapping.  
It can be applied to many problems.  For example, 
the cluster formalism does not work well for frustrated 
systems, but we can use the generalized scheme of the PCC algorithm.  
We can also apply the generalized scheme 
to a wide variety of quantum systems. 

We thank N. Kawashima, K. Harada, H. Otsuka, K. Nomura, 
S. Todo, C. Yasuda, Y. Ozeki, and F. Alet for fruitful discussions.  
This work was supported by a Grant-in-Aid for Scientific Research 
from the Japan Society for the Promotion of Science.

\end{document}